\documentclass[12pt]{article}
\usepackage{amsfonts,amsmath}
\usepackage{amssymb}

\renewcommand{\theequation}{\thesection.\arabic{equation}}
\renewcommand{\thesubsection}{\arabic{section}.\arabic{subsection}}
\makeatletter \@addtoreset{equation}{section} \makeatother

\def\abz{}
\def\vac{F_K}

\def\al{\alpha}

\def\*{\star}

\newcommand{\be}{\begin{equation}}
\newcommand{\ee}{\end{equation}}
\newcommand{\bee}{\begin{eqnarray}}
\newcommand{\beee}{\begin{array}}
\newcommand{\eee}{\end{eqnarray}}
\newcommand{\eeee}{\end{array}}
\newcommand{\ga}{\alpha}
\newcommand{\gb}{\beta}
\newcommand{\gga}{\gamma}

\newcommand{\D}{{\cal D}}
\newcommand{\Q}{{\cal Q}}
\newcommand{\K}{{\cal K}}

\newcommand{\F}{{\mathbf f}_{0}}
\newcommand{\ls}{\!\!\!\!\!\!}

\newcommand{\gd}{\delta}
\newcommand{\gl}{\lambda}
\newcommand{\gk}{\varkappa}
\newcommand{\gep}{\epsilon}

\newcommand{\gs}{\sigma}
\newcommand{\go}{\omega}
\newcommand{\gO}{\Omega}

\newcommand{\dal}{\dot \alpha}
\newcommand{\dgb}{\dot \beta}
\newcommand{\dgga}{\dot \gamma}

\newcommand{\q}{\,,\qquad}

\newcommand{\half}{\frac{1}{2}}

\newcommand{\p}{\partial}

\newcommand{\ff}{\frac}

\renewenvironment{thebibliography}[1]
 { {\bf \LARGE References} \begin{list}{[\arabic{enumi}]}
  {\usecounter{enumi} \setlength{\parsep}{2pt}
  \setlength{\itemsep}{0pt} \settowidth{\labelwidth}{#1.}
   \sloppy
  }}{\end{list}}

\begin{document}

\begin{flushright}
FIAN/TD/10-09\\
\end{flushright}

\vspace{0.5cm}
\begin{center}
{\large\bf Static BPS black hole in $4d$ higher-spin gauge theory}

\vspace{1 cm}

{\bf V.E.~Didenko and M.A.~Vasiliev}\\
\vspace{0.5 cm} {\it
I.E. Tamm Department of Theoretical Physics, Lebedev Physical Institute,\\
Leninsky prospect 53, 119991, Moscow, Russia }

\vspace{0.6 cm}
didenko@lpi.ru, vasiliev@lpi.ru \\
\end{center}

\vspace{0.4 cm}

\begin{abstract}
\noindent We find exact spherically symmetric solution of $4d$ nonlinear
bosonic higher-spin gauge theory, that
preserves a quarter of supersymmetries of
$\mathcal{N}=2$ supersymmetric $4d$ higher-spin gauge theory. In
the weak field regime it describes $AdS_4$
Schwarzschild black hole in the spin two sector along with
non-zero massless fields of all integer spins.
\end{abstract}

\section{Introduction}

Nonlinear field equations of  higher spin (HS) gauge fields were
originally found for the $4d$ HS theory  \cite{Vas4,
Vas4-more}. Then these results were extended to $d=3$ \cite{Vas3}
and any $d$ \cite{Vasd}. In $d\geq 4$, HS gauge
theories describe interactions of infinite sets of propagating
massless fields of lower and higher spins. The theory is
manifestly general coordinate invariant, containing gravity as its
part.  In HS theory, spin two sources HS fields and vice versa. As
a result, solutions of Einstein gravity are not necessarily
solutions of the HS theory. HS interactions may significantly
affect the theory in the strong field regime. Moreover, since the
interval $ds^2=g_{mn}dx^m dx^n$ associated to the $s=2$ field
$g_{mn}$ is not invariant under HS gauge symmetry, standard
concepts of general relativity have to be reconsidered in the HS
theory with unbroken HS symmetries.

\abz The difficulty of finding solutions in HS theory is due to
nonlocality of the field equations in the auxiliary noncommutative
twistor space, rendered by the star-product operation. By field
equations this nonlocality in the  twistor space is mapped to
space-time nonlocality of the HS field equations at the
interaction level in accordance with the fact that HS interactions
contain higher derivatives \cite{BBB,BBD}. (For review and more
references on HS gauge theories see \cite{solv}.)  As a result,
 the nonlinear HS gauge theory goes beyond
the low energy limit typical for Einstein theory and its
perturbative stringy corrections, having capacity to account
strong field effects.

\abz So far a very few exact solutions of the HS theory were
available. The simplest one is the $AdS_d$ vacuum solution.
BTZ black hole (BH) \cite{BTZ} also solves $3d$ nonlinear
HS field equations \cite{Didenko:2006zd}. The first
nontrivial example of exact solution of $3d$ HS theory was found
in \cite{ProkVas}, where it was shown that for non-zero HS curvature
vacuum field $B_0=\nu$, the $3d$ HS field equations describe
massive matter fields with the $\nu$-dependent mass parameter. Some
solutions of $4d$ HS equations were
studied by Sezgin and Sundell  \cite{SS4d,SS-exact}. More general
class of solutions with non-vanishing HS fields, which reveal
examples of algebraically special solutions of HS theory, was
obtained by Iazeolla, Sezgin and Sundell in \cite{Ia}.
 A large class of solutions of this type solves
 chiral models with $(0,4)$ or $(2,2)$
space-time signatures.  In the noncommutative twistor sector
they effectively amount to the $3d$ solution of \cite{ProkVas},
but their physical meaning  is still not completely clear.


\abz Since  HS theory extends  Einstein gravity, the natural
question is what are  counterparts of BH solutions in HS theory?
In this paper we answer this question for the simplest case of a
spherically symmetric charged BH. Namely, we present explicit
coordinate free construction of the exact solution of the bosonic
sector of $\mathcal{N}=2$ supersymmetric $4d$ HS theory of
\cite{Vas4-more}, characterized by a single free dimensionful
parameter $M$ (for a given cosmological constant) that
determines a value of the BH electric charge $e$. In the weak
field approximation, where the terms proportional to $e^2$ are
neglected, the obtained solution reproduces  Schwarzschild BH of
mass $M$ for $s=2$ along with BH spin-$s$ massless fields found in
\cite{DMV1}. Beyond this limit, the obtained solution differs from
Schwarzschild BH being supersymmetric, {\it i.e.}, BPS. This in
turn suggests that its electric charge is critical and hence
nonzero.

Our construction is based on the unfolded formulation of Einstein
BH developed in \cite{DMV2} where $AdS_4$ Killing
symmetries play a distinguished role. The presented HS BH solution
is constructed in terms of a  Fock vacuum in the HS star--product algebra
associated to a particular Killing symmetry of $AdS_4$. This Ansatz
exhibits dramatic simplification of the
perturbative analysis of the nonlinear HS equations,
reducing the problem to $3d$ HS equations. The same time, it
 makes the obtained solution supersymmetric.

\abz The paper is organized as follows. In Section \ref{BH-unfo}
we recall the formulation of $AdS_4$ BH of \cite{DMV2}. Nonlinear
$4d$ HS bosonic equations are recalled in Section \ref{4d-eqs}. In
Section \ref{BH-free HS} we explain how BH and its HS
extension arises in the free HS theory. The exact HS BH
solution of  the nonlinear HS field equations is obtained in Section
\ref{BH-nonlin}. Its symmetries are found in Section
\ref{global}. Conclusions and perspectives are discussed in
Section \ref{conc}. Conventions are summarized in Appendix.

\section{$AdS_4$ Schwarzschild black hole}\label{BH-unfo}

As is well-known, the metric of a BH of mass $M$ in $AdS_4$ admits
the Kerr-Schild form\footnote{In this paper we set Newton constant $4\pi
G=1$.} \cite{Carter}
\be\label{KS-metric}
g_{mn}=\eta_{mn}+\ff{2M}{r}k_{m}k_{n}\q
g^{mn}=\eta^{mn}-\ff{2M}{r}k^{m}k^{n}\,,
\ee
where $\eta_{mn}$, $(m,n=0\dots 3)$ is the background $AdS_4$
metric with negative cosmological constant $-\gl^2$, $k^{m}$ is
the Kerr-Schild vector, that  satisfies \be\label{KS-cond}
k^{m}k_{m}=0\,,\quad k^{m}\D_{m}k_{n}=k^{m}D_{m}k_{n}=0\,, \ee
where $\D_{m}$ and $D_m$ are BH and $AdS_4$ covariant derivatives,
respectively, and
\be\label{r} \ff1r=-\ff12\D_{m}k^m=-\ff12D_{m}k^m\,. \ee Note,
that it makes no difference  in \eqref{KS-metric}-\eqref{r}
whether indices are raised and lowered by either $AdS_4$ or full
BH metrics. Although in this paper we proceed with the coordinate
independent description, let us present explicit decomposition
\eqref{KS-metric} in a coordinate system of \cite{Gibon} where
$k^m$ has a particularly simple form \be\label{fon} \eta_{mn}=
\ff{1}{1+\gl^2r^2}\left(
\begin{array}{cccc}
  (1+\gl^2r^2)^2 & 0 & 0 & 0 \\
  0 & -1-\lambda^2(y^2+z^2) & \lambda^2 xy &  \lambda^2 xz\\
  0 & \lambda^2 xy & -1-\lambda^2(x^2+z^2) & \lambda^2 yz\\
  0 & \lambda^2 xz & \lambda^2 yz & -1-\lambda^2(x^2+y^2)\\
\end{array}
\right)
\ee
\be
k^0=\ff{1}{1+\gl^2r^2}\,,\quad k^1=-\ff{x}{r}\,,\quad
k^2=-\ff{y}{r}\,,\quad k^3=-\ff{z}{r}\,,\quad r^2=x^2+y^2+z^2\,.
\ee

\abz
Let $AdS_4$ be described by the zero-curvature equation
\be\label{zero}
R_{0AB}=dW_{0AB}+\half W_{0A}{}^{C}\wedge W_{0CB}=0\q
A,B=1\dots 4\,,
\ee
where $d=dx^n \frac{\p}{\p x^n}$ and $ W_{0AB}(x)=W_{0BA}(x)$ is the
$sp(4)$ connection 1-form.
Indices $A,B,\dots$ are raised and lowered by the
$sp(4)$ symplectic form identified with $4d$ charge conjugation
matrix (see Appendix). In two-component spinor
notation
\be
W_{0AB}=\left(
\begin{array}{cc}
\go_{\al\gb} & -\gl h_{\al\dgb}\\
-\gl h_{\gb\dal} & \bar{\go}_{\dal\dgb}\\
\end{array}
\right)\,,\quad \go_{\al\gb}=\go_{\gb\al}\,,\quad
\bar{\go}_{\dal\dgb}=\bar{\go}_{\dgb\dal}\,,
\ee
where 1-forms $\go_{\al\gb}$ and $h_{\al\dal}$ describe the
Lorentz connection and vierbein, respectively.

\abz $AdS_4$ possesses ten global symmetry parameters
$\mathcal{K}_{AB}=\mathcal{K}_{BA}$ valued in the $o(3,2)\sim
sp(4)$ Lie algebra, that satisfy
\be\label{glob}
D_{0}\mathcal{K}_{AB}=0\q D_0^2=0\,,
\ee
where $D_0$ is the $AdS_4$ covariant differential (e.g., $D_0 A_A
= dA_A +\ff12W_{0\,A}{}^B A_B$) that squares to zero by virtue of
(\ref{zero}). In terms of two-component spinors with
\be\label{param}
\mathcal{K}_{AB}=\left(
\begin{array}{cc}
\gl^{-1}\gk_{\al\beta} & v_{\al\dgb}\\
v_{\gb\dal} & \gl^{-1}\bar{\gk}_{\dal\dgb}\\
\end{array}\right)\,,
\ee
\eqref{glob} amounts to
\be\label{Kill}
D^{L}v_{\al\dal}=\half h^{\gga}{}_{\dal}\gk_{\gga\al}+\half
h_{\al}{}^{\dgga}\bar{\gk}_{\dal\dgga}\q
D^{L}\gk_{\al\al}=\gl^2h_{\al}{}^{\dgga}v_{\al\dgga}\q
D^{L}\bar{\gk}_{\dal\dal}=\gl^2h^{\gga}{}_{\dal}v_{\gga\dal}\,,
\ee
where
\[
D^{L}A_{\al}=dA_{\al}+\half\go_{\al}{}^{\gga}A_{\gga}\q
D^{L}\bar{A}_{\dal}=d\bar{A}_{\dal}+\half\bar{\go}_{\dal}{}^{\dgga}\bar{A}_{\dgga}\,.
\]
{}From (\ref{Kill}) it follows in particular that $v_{\al\dal}$
is  an $AdS_4$ Killing vector in the local frame.

\abz According to \cite{DMV2}, the Kerr--Schild vector of the BH
metric \eqref{KS-metric} and its derivatives are expressed in
terms of $\mathcal{K}_{AB}$ \eqref{param}
\be\label{KS}
k_{\al\dal}\equiv
h^{n}_{\al\dal}k_{n}=\ff{1}{v^-v^+}v^{-}_{\al\dal}\q
v^{\pm}_{\al\dal}=\pi^{\pm}_{\al}{}^{\gga}\bar{\pi}^{\pm}_{\dal}{}^{\dgga}v_{\gga\dgga}\q
v^-v^+=\half v^-_{\al\dal}v^{+\al\dal}\,,
\ee
where
\be\label{proj}
\pi^{\pm}_{\al\gb}=\ff12(\gep_{\al\gb}\pm\ff{\gk_{\al\gb}}{\sqrt{-\gk^2}})\q
\gk^2=\ff12\gk_{\al\gb}\gk^{\al\gb}\,
\ee
are projectors, that satisfy
\be\label{pi-prop}
\pi^{\pm}_{\al}{}^{\gga}\pi^{\pm}_{\gga\gb}=\pi^{\pm}_{\al\gb}\q
\pi^{\pm}_{\al}{}^{\gga}\pi^{\mp}_{\gga\gb}=0\,.
\ee
(The conjugated projectors $\bar{\pi}^{\pm}_{\dot{\gamma}\dgb}$ satisfy analogous
relations.)
Note that, alternatively, one can use  $v^{+}_{\al\dal}$ in
\eqref{KS} instead of $v^{-}_{\al\dal}$ \cite{DMV1}.

\abz A type of the BH (e.g., whether it is rotating or static)
depends on the values of $sp(4)$ invariants associated to
$\mathcal{K}^{AB}$. The static case is characterized by the
condition
\be\label{schwarz}
\mathcal{K}_{A}{}^{B}\mathcal{K}_{B}{}^{C}=-\gd_{A}{}^{C}\,
\ee
equivalent to
\be\label{sch-1}
\gl^{-2}\gk^2+v^2=1\,,\quad \gk^2=\bar{\gk}^2\,,\quad
\bar{\gk}_{\dal}{}^{\dgga}v_{\gb\dgga}+v^{\gga}{}_{\dal}\gk_{\gga\gb}=0\,.
\ee
The function $r$ \eqref{r} satisfies \cite{DMV2}
$$
\ff1r=\ff{\gl^2}{\sqrt{-\gk^2}}\q
d\left (\ff1r\right)=\ff{1}{2\gl^{2}r^3}h^{\al\dal}v^{\al}{}_{\dal}\gk_{\al\al}\,.
$$
Einstein equations for the metric \eqref{KS-metric} are satisfied as a
consequence of \eqref{glob} and \eqref{KS}.

\abz The Kerr-Schild BH solution admits a
{\it HS Kerr-Schild} generalization for
massless bosonic fields of all spins \cite{DMV1}
\be\label{fronsd}
\phi_{m_1\dots m_k}=\ff{2M}{r}k_{m_1}\dots k_{m_k}
\ee
that satisfy free spin-$s$ equation in the $AdS_4$ background
\be\label{free-s}
D^n D_n \varphi_{m(s)}-sD_n D_m
\varphi^{n}{}_{m(s-1)}=-2(s-1)(s+1)\gl^2\varphi_{m(s)}\,.
\ee
The case of $s=2$ reproduces Kerr-Schild term of the BH
metric \eqref{KS-metric}. As shown in Sections \ref{BH-free HS} and
\ref{BH-nonlin}, the  HS Kerr-Schild fields solve free field
equations of the linearized HS theory and remain nonzero in
the obtained HS BH solution.

\section{HS equations in four dimensions}\label{4d-eqs}

To reproduce Kerr-Schild solution in HS theory let us recall the
structure of nonlinear $4d$ HS equations. (For more detail see
\cite{Vas-rev}.)
Starting from this section we set $\gl^2=1$ for convenience.

\abz HS nonlinear equations in $d=4$ are formulated in terms of
1-forms $ W(Z,Y|x)= dx^n W_n(Z,Y|x) $ and 0-forms $B(Z,Y|x)$ that
depend on space-time coordinates $x^n$ and auxiliary spinor
variables $Z^A$ and $Y^A$. In addition, the 1-form connection
along $Z$-direction should be introduced
$S(Z,Y|x)=S_{\al}(Z,Y|x)dz^{\al}+\bar{S}_{\dal}(Z,Y|x)d\bar{z}^{\dal}
$ to be expressed via dynamical fields by the field equations.
It is required that $\{dx^n, dz^\al\}=0$, $ \{dx^n,
d\bar{z}^{\dal}\}=0$. In this section, we consider the bosonic sector
of the HS equations of \cite{Vas4-more}, where  the fields $B$ and
$W$ are even functions of the oscillators $(Z,Y)$ and $S$ is odd.
The simplest version of HS equations with the topological fields
factored out is
\be\label{DW}
dW-W\*\wedge W=0\,,
\ee
\be\label{DB}
dB-W\*B+B\*\tilde{W}=0\,,
\ee
\be\label{DS}
dS_{\al}-[W,S_{\al}]_{\*}=0\,,\quad
d\bar{S}_{\dal}-[W,\bar{S}_{\dal}]_{\*}=0\,,
\ee
\be\label{SS}
S_{\al}\*S^{\al}=2(1+B\*v)\,,\quad \bar{S}_{\dal}
\*\bar{S}^{\dal}=2(1+B\*\bar{v})\,,\quad [S_{\al},
\bar{S}_{\dal}]_{\*}=0\,,
\ee
\be\label{BS}
B\*\tilde{S}_\al+S_\al\*B=0\,,\quad
B\*\tilde{\bar{S}}_{\dal}+\bar{S}_{\dal}\*B=0\,,
\ee
where $ \tilde{A}=(-u_\al, \bar{u}_{\dal})$ for $A=(u_\al,
\bar{u}_{\dal})$ and
\be\label{Klein}
v=\exp{(z_\al y^\al)}\q
\bar{v}=\exp{(\bar{z}_{\dal}\bar{y}^{\dal})}\,.
\ee

The star-product in the auxiliary space of commuting variables
$Y_A=(y_\al, \bar{y}_{\dal})$, $Z_A=(z_\al, \bar{z}_{\dal})$
is defined by
\be\label{star}
(f\star g)(Z,Y)=\ff{1}{(2\pi)^8}\int d^4ud^4vf(Z+U, Y+U)g(Z-V,
Y+V)e^{U_{A}V^{A}}\,,\quad U_AV^A=u_\al
v^\al+\bar{u}_{\dal}\bar{v}^{\dal}\,.
\ee
An integration contour in \eqref{star} is chosen so that $1\*
f=f\*1=f$. Note that the star-product definition (\ref{star})
differs from that of \cite{Vas-rev} by the absence of the imaginary
unit factor in the exponential. {}From (\ref{star}) it follows in particular
\[
Y_A\star f=(Y_A-\ff{\p}{\p Y^A}+\ff{\p}{\p Z^A})f\,,\quad f\star
Y_A=(Y_A+\ff{\p}{\p Y^A}+\ff{\p}{\p Z^A})f\,,
\]
\be\label{star-prop}
Z_A\star f=(Z_A-\ff{\p}{\p Y^A}+\ff{\p}{\p Z^A})f\,,\quad f\star
Z_A=(Z_A-\ff{\p}{\p Y^A}-\ff{\p}{\p Z^A})f\,,
\ee
\be
[z_\al, z_\gb]_\*=-[y_\al, y_\gb]_\*=2\gep_{\al\gb}\,,\quad
[\bar{z}_{\dal}, \bar{z}_{\dgb}]_\*=-[\bar{y}_{\dal},
\bar{y}_{\dgb}]_\*=2\gep_{\dal\dgb}\,,\quad [y_{\al},
\bar{y}_{\dal}]_\* =[z_{\al}, \bar{z}_{\dal}]_\*=0\,.
\ee
An important property of the star-product \eqref{star} is that it
admits left and right inner Klein operators (\ref{Klein}) that satisfy
\be\label{v-pr1}
v\*v=\bar{v}\*\bar{v}=1\q v\*f(z,y)=f(-z,-y)\*v\q
\bar{v}\*f(\bar{z}, \bar{y})=f(-\bar{z}, -\bar{y})\*\bar{v}
\ee
and
\be\label{v-pr2}
v\*f(z,y)=\exp{(z_\al y^\al)}f(y,z)\q \bar{v}\*f(\bar{z},
\bar{y})=\exp{(\bar{z}_{\dal}\bar{y}^{\dal})}f(\bar{y},\bar{z})\,.
\ee
Note, that  Klein operators, that act only on $Y$ or $Z$ variables,
are delta-functions \cite{Ber-Schub} (hence not entire)
\be\label{del-2}
\gk_y\*\gk_y=1\q \gk_y\*f(y)=f(-y)\*\gk_y\,,\quad
\gk_y=2\pi\gd^{(2)}(y)
\ee
Also note that $\hat{f}(y)=f(y)\*\gk_y$ gives  Fourier transform
with respect to $y$--variable $ \hat f(y)=\int d^2u f(u)e^{-u_\al
y^\al}\,. $ Analogous formulae hold for dotted spinors and for
$y\leftrightarrow z$.

 It follows then that
\be\label{del-1}
v=\gk_y\*\gk_z\q \bar{v}=\gk_{\bar y}\*\gk_{\bar{z}}\,.
\ee
That the results are entire functions rather than  distributions
is because the star-product (\ref{star}) is a specific normal
star-product rather than Weyl star-product (for more detail see
\cite{Vas-rev}).

The Eqs. \eqref{DW}-\eqref{BS} are consistent and
manifestly invariant under the gauge transformations
\be\label{gauge-B}
\gd B=\gep\*B-B\*\tilde{\gep}\,,\quad \gd W=d\gep+[\gep,
W]_{\*}\,,\quad \gd S_\al=[\gep, S_\al]_{\*}\,,\quad \gd
\bar{S}_{\dal}=[\gep, \bar{S}_{\dal}]_{\*}
\ee
with an arbitrary gauge parameter $\gep=\gep(Z, Y|x)$. The vacuum
solution for \eqref{DW}-\eqref{BS}, that describes empty $AdS_4$
space, is $ B_0=0\,,\quad S_0=z_\al
dz^{\al}+\bar{z}_{\dal}d\bar{z}^{\dal}\,,\quad W_0=W_0(Y|x)\,,$
where
\be\label{W0}
W_0(Y|x)=-\ff18\Big(\go_{\al\al}(x)y^\al
y^\al+\bar{\go}_{\dal\dal}(x)\bar{y}^{\dal}\bar{y}^{\dal}-
2h_{\al\dal}(x)y^\al\bar{y}^{\dal}\Big)
\ee
describes $AdS_4$ vacuum fields via Eq.~(\ref{zero}) that acquires the form
\be\label{zero-W}
\D_0^2\equiv dW_0-W_0\*\wedge W_0=0\,.
\ee

\abz The variables $Z_A$ in the star-product \eqref{star}, that
play  important role in the description of HS interactions,
can be neglected at the
free level. Following \cite{Vas-1st order,Vas-rev} a free HS field
is described in the unfolded formalism by the 1-forms $w(Y|x)$ and
0-forms $C(Y|x)$
\be
\ls w(Y|x)=\sum_{n,m=0}^{\infty}
\ff{1}{n!m!}w_{\al(n),\dal(m)}y^\al\dots y^\al \bar{y}^{\dal}\dots
\bar{y}^{\dal},\quad C(Y|x)=\sum_{n,m=0}^{\infty}
\ff{1}{n!m!}C_{\al(n),\dal(m)}y^\al\dots y^\al \bar{y}^{\dal}\dots
\bar{y}^{\dal}
\ee
that encode, respectively, linearized HS potentials and field
strengths along with towers of auxiliary fields. $w(Y|x)$ and
$C(Y|x)$ are the parts of
$W(Z,Y|x)$ and $B(Z,Y|x)$,
 that remain unrestricted by the equations (\ref{DS})-(\ref{BS}).
In the sector of 0-forms, free equations resulting from (\ref{DB})
have a form of covariant constancy condition in the
twisted--adjoint module
\be\label{C-mod}
\tilde{\D_0}C\equiv dC-W_0\*C+C\*\tilde{W}_0=0\q
\tilde{f}(y,\bar{y})=f(-y,\bar{y})\,.
\ee
 The linearized equations for HS gauge potentials, that result
from (\ref{DW}),
 are \cite{Vas-1st order}
\be\label{R1}
R_{1\al(n),\dal(m)}=\gd(m)h^{\gga\dgb}\wedge
h^{\gga}{}_{\dgb}C_{\al(n)\gga(2)}+\gd(n)h^{\gga\dgb}\wedge
h_{\gga}{}^{\dgb}C_{\dal(m)\dgb(2)}\,
\ee
($\delta(n)= 1(0)$ at $n=0(n\neq 0)$),
where the curvatures $R_{1\al(n),\dal(m)}$ are components of the
linearized HS curvature tensor
\be\label{D0w}
R_1(Y|x)\equiv \D_0w(Y|x)\,.
\ee
Note, that Eq.~\eqref{C-mod} with a chosen $W_0$ is invariant under HS
global symmetry transformation
\be\label{C-glob}
\gd C=\gep_0\* C-C\*\tilde{\gep}_0\,
\ee
provided that
\be\label{HS-glob}
\D_0\gep_0=0\,.
\ee
Since the twist operation in \eqref{C-mod} can be  realized as $
\tilde{f}(Y)=\gk_y\*f\*\gk_y\,, $ any solution of the global
symmetry equation \eqref{HS-glob} $\gep_0(Y|x)$ generates a
solution of \eqref{C-mod} of the form
\be\label{C-sol}
C(Y|x)=c_1\gep_0(Y|x)\*\gd(y)+c_2\gep_0(Y|x)\*\gd(\bar{y})\,,
\ee
where $c_1, c_2$ are arbitrary constants. This formula manifests
that adjoint and twisted adjoint covariant derivatives are related
via Fourier transform of either $y_\al$ or $ \bar{y}_{\dal}$ variables.

\section{Black hole solution in free HS theory}\label{BH-free HS}

\abz As shown in \cite{DMV2}, a generic $AdS_4$ BH is completely
determined by a chosen global symmetry parameter
$\mathcal{K}_{AB}$ of $AdS_4$. Let us apply this idea to a HS
generalization. Since $\mathcal{K}_{AB}$ satisfies \eqref{glob},
it follows that
\be\label{DK}
\D_0 f(\mathcal{K}_{AB}Y^AY^B)=0\,
\ee
for any $f(\xi)$. By \eqref{C-sol}, a
solution of the free HS equation \eqref{C-mod}
is generated by
\be\label{NUT}
C(Y|x)=Mf(\mathcal{K}_{AB}Y^A Y^B)\*\gk_y\,.
\ee
 Generally, $C$ \eqref{NUT} is not Hermitian yielding
at the linearized level two different real solutions. In the $s=2$
sector each can be shown to correspond  to generic
$AdS_4$--Kerr--NUT BH with mass $m\sim \text{Re } M$ and NUT
charge $n\sim \text{Im } M$  for one
and vice versa for another.

In this paper we confine ourselves to the simplest static case of
\eqref{schwarz} with real $M$ and $f (\xi)=4\exp(i\xi/2)$,
\be\label{exp}
F_K=4\exp{(\ff{i}{2}\mathcal{K}_{AB}Y^A Y^B)}\,.
\ee
To avoid an extra $i$ in exponential \eqref{exp} it is convenient
to introduce
\[
K_{AB}=i\mathcal{K}_{AB}\,,\quad
K_{A}{}^{C}K_{C}{}^{B}=+\gd_{A}{}^{B}\,.
\]
The coefficient $1/2$ in the exponential \eqref{exp} is chosen so
that, due to \eqref{schwarz}, $F_K$ satisfies
\be\label{id}
\vac\*\vac=\vac\q F_K\*\gk_y=F_K\*\gk_{\bar y}\,.
\ee
The first order HS BH curvature $C(Y|x)=MF_K\*\gk_y$ has the form
\be\label{C-schwarz}
C(Y|x)=\ff{4M}{r}\exp{\Big(\ff12\gk^{-1}_{\al\gb}y^{\al}y^{\gb}+
\ff12\bar{\gk}^{-1}_{\dal\dgb}\bar{y}^{\dal}\bar{y}^{\dgb}+
\gk^{-1}_{\al\gga}v^{\gga}{}_{\dal}y^{\al}\bar{y}^{\dal}\Big)}\,,
\ee
where $\gk^{-1}_{\al\gb}=-\ff{1}{\gk^2}\gk_{\al\gb}$ and
$r=\sqrt{-\gk^2}$. From \eqref{C-schwarz} it follows that HS Weyl
tensors are
\be\label{schwarz-Weyl}
C_{\al(2s)}=\ff{M}{2^{s-2} r}(\gk^{-1}_{\al\al})^s\q
\bar{C}_{\dal(2s)}=\ff{M}{2^{s-2}
r}(\bar{\gk}^{-1}_{\dal\dal})^s\,.
\ee
According to \cite{DMV2}, these are Petrov type--D Weyl tensors,
that describe a Schwarzschild BH of mass $M$ in the $s=2$ sector
 along with a tower of Kerr-Schild fields (\ref{fronsd}) of all spins $s$.
 Note that the
frame-like HS connections corresponding to \eqref{schwarz-Weyl}, that carry equal numbers of undotted
and dotted spinors, can be shown to be gauge equivalent to
\be\label{Fr}
W_{phys}=
\ff{4M}{r}h^{\al\dal}k_{\al\dal}\exp{\Big(-\ff12k_{\gb\dgb}y^{\gb}\bar{y}^{\dgb}\Big)}\,.
\ee

\section{Black hole in nonlinear HS theory}\label{BH-nonlin}
Starting with Schwarzschild solution \eqref{C-schwarz} at the free level,
we have to analyze higher-order corrections within the system
(\ref{DW})-(\ref{BS}). The specific choice \eqref{exp} simplifies
 the analysis, allowing
us to solve the problem exactly.

\abz To reconstruct the HS  field strength $B(Z,Y|x)$ and the
1-form connection $S(Z,Y|x)$ along $Z$ directions one has first to
solve the constraints \eqref{SS}, \eqref{BS} which form a closed
subsystem. This fixes dynamics completely reducing the problem to
determination of HS potentials in terms of their curvatures via
the equations \eqref{DW}-\eqref{DS}.

\subsection{BH Fock vacua}

The key fact is that $F_K$ \eqref{exp} generates a Fock vacuum of the
star-product
algebra, defining a subalgebra of a reduced set of oscillators.
Indeed, let us introduce the projectors $\Pi_{\pm AB}$
\be\label{Pi}
\Pi_{\pm AB}=\ff{1}{2}(\gep_{AB}\pm K_{AB})\q
\Pi_{\pm A}{}^{C}\Pi_{\pm C}{}^{B}=\Pi_{\pm A}{}^{B}\,,\quad
\Pi_{\pm A}{}^{C}\Pi_{\mp C}{}^{B}=0\,.
\ee
The creation and annihilation operators $ Y_{\pm A}=\Pi_{\pm
A}{}^{B}Y_B $ satisfy
\be
[Y_{+A}, Y_{+B}]_\*=[Y_{-A}, Y_{-B}]_\*=0\q [Y_{+A},
Y_{-B}]_\*=\Pi_{+AB}\,,
\ee
\be\label{fock}
Y_{-A}\*\vac=\vac\*Y_{+A}=0\,.
\ee

 In the nonlinear case,  we have to analyze
corrections due to $Z$--dependence of the HS curvature and
connection. $F_K$ is the only
$Z$--independent element of the star-product algebra (\ref{star})
 that satisfies \eqref{fock}. More generally, let $F$ be a space of elements
$f(Z,Y|x)$ that satisfy
\be\label{def}
Y_{-A}\*f=f\*Y_{+A}=0\,:\qquad f=\vac\phi(A|x)\,,
\ee
where $\phi$ is an arbitrary function and
\be\label{A-def}
A_{A}=(a_\al,
\bar{a}_{\dal})=Y_{+A}+Z_{+A}-(Y_{-A}-Z_{-A})=Z_A+K_{A}{}^{B}Y_B\q
[A_A, A_B]_\*=4\gep_{AB}\,.
\ee
$F$ is a subalgebra of the star--product algebra.
Namely,
\be\label{comp}
\Big(\vac\phi_1\Big)\*\Big(\vac\phi_2\Big)=\vac(\phi_1*\phi_2)\,,
\ee
where we have introduced the induced star--product  $*$
on the space of functions $\phi(A|x)$, that has the form
\be\label{in-star}
(\phi_1*\phi_2)(A)=\int
d^4U\phi_1(A+2U_+)\phi_2(A-2U_-)e^{2U_{+A}U_{-}^A}\,.
\ee
The integral \eqref{in-star} is normalized in such a way that
$1*\phi=\phi*1=\phi$. Star--product \eqref{in-star} is associative
and describes  the normal ordering of
the operators $(Y_{-A}-Z_{-A})$ and $(Y_{+A}+Z_{+A})$.

\abz Note that any function of the form
$\tilde{F}_K=f(Z|x)\*\vac$ satisfies \eqref{fock} and, hence, can
be represented in the form \eqref{def}. Indeed, using \eqref{star},
one can easily check  that
\be
\label{za}
\vac\*f(Z|x)=\ff14\vac\int d^2 v d^2\bar v
f(A-V|x)e^{\ff12K_{AB}V^{A}V^{B}}\,.
\ee
For (anti)holomorphic functions, the integration can be performed
further
\be
\vac\* f(z)=\ff{1}{2r}\vac\int d^2 v f(a-v)e^{\ff12
\gk^{-1}_{\al\gb}v^\al v^{\gb}}\,,\quad  \vac\*
f(\bar{z})=\ff{1}{2r}\vac\int d^2\bar v f(\bar a-\bar v)e^{\ff12
\bar{\gk}^{-1}_{\dal\dgb}\bar{v}^{\dal} \bar{v}^{\dgb}}\,.
\ee
In the sequel we will work with (anti)holomorphic functions
of  ($\bar{a}_{\dal}$)$a_{\al}$, using relations
\be\label{a-com}
[a_{\al}, f(a)]_*=2\ff{\p}{\p a^{\al}}f(a)\q \{a_{\al},
f(a)\}_*=2(a_\al+\gk_{\al}{}^{\gb}\ff{\p}{\p a^{\gb}})f(a)\q
[a_\al, \bar{a}_{\dal}]_*=0\,.
\ee

\abz The star--product \eqref{in-star} possesses the Klein
operators $\K$ and $\bar \K$,
\be\label{Klein1}
\K=\ff1r\exp{\Big(\ff12\gk^{-1}_{\al\al}a^\al a^\al\Big)}\q
\bar{\K}=\ff1r\exp{\Big(\ff12\bar{\gk}^{-1}_{\dal\dal}\bar{a}^{\dal}
\bar{a}^{\dal}\Big)}
\ee
that satisfy
\be\label{Klein-prop}
\K*\K=\bar \K*\bar \K=1\,,\quad \{\K,a_\al\}_*=\{\bar
\K,\bar{a}_{\dal}\}_*=0\,,\quad [\K,\bar
\K]_*=[\K,\bar{a}_{\dal}]_*=[\bar \K, a_{\al}]_*=0\,
\ee
and  result from
\be\label{prop2}
\vac\*\gk_z=\vac \K\q \vac\*\gk_{\bar z}=\vac\bar \K\,.
\ee

\subsection{Ansatz and final result}
The key observation is that the Eqs.~\eqref{SS},  \eqref{BS} can be
solved exactly by the Ansatz
\be\label{B}
B=M\vac\*\gk_y\,,
\ee
\be
S_\al=z_\al+\vac\gs_{\al}(a|x)\q
\bar{S}_{\dal}=\bar{z}_{\dal}+\vac\bar{\gs}_{\dal}(\bar{a}|x)
\ee
with some functions $\gs_{\al}(a|x)$,
$\bar{\gs}_{\dal}(\bar{a}|x)$ to be specified later.
Within the induced star--product \eqref{in-star}, this
Ansatz reduces \eqref{SS}, \eqref{BS}  to two copies of $3d$
(anti)holomorphic {\it deformed oscillators} considered in \cite{Aq,ProkVas}
in the context of $3d$ HS theories. Indeed, introducing
\be
s_{\al}=a_{\al}+\gs_{\al}(a|x)\q
\bar{s}_{\dal}=\bar{a}_{\dal}+\bar{\gs}_{\dal}(\bar{a}|x)
\ee
and using \eqref{del-1}, \eqref{id}, \eqref{a-com},
\eqref{Klein-prop} and \eqref{prop2} one arrives at the following
system
\be\label{deform}
[s_{\al}, s_{\gb}]_*=2\gep_{\al\gb}(1+M \K)\q \{\K,s_\al\}_*=0\q
\K*\K=1\,.
\ee
The deformed oscillators \eqref{deform} were originally discovered
by Wigner in \cite{Wigner} in a somewhat different form. The BH
mass $M$ plays a role of the deformation parameter.

\abz Analogous equations hold in the dotted sector
\be\label{deform-dot}
[\bar{s}_{\dal}, \bar{s}_{\dgb}]_*=2\gep_{\dal\dgb}(1+M\bar{\K})\q
\{\bar{s}_{\dal}, \bar{\K}\}_*=0\q \bar \K*\bar \K=1\,.
\ee
In addition,
\be\label{deform-mix}
[s_\al, \bar{s}_{\dal}]_*=0\,,\quad [s_\al, \bar{\K}]_*=0\,,\quad
[\bar{s}_{\dal}, \K]_*=0\,,\quad [\K,\bar \K]_*=0\,.
\ee
Eqs. \eqref{deform}-\eqref{deform-mix} follow
from \eqref{Klein-prop} and \eqref{a-com}.

\abz A proper Ansatz for HS connection 1-form $W(Y,Z|x)$ is
\be\label{con}
W=W_0+\vac\Big(\go(a|x)+\bar{\go}(\bar{a}|x)\Big)\,,
\ee
where $W_0(Y|x)$ is the vacuum connection \eqref{W0} and the additional terms manifest
holomorphic factorization  with respect to the
variables ($a_{\al}, \bar{a}_{\dal}$). Note, that
\eqref{con} has no definite holomorphy properties in the
($y_\ga, \bar{y}_{\dot{\ga}}$) variables because both $a_{\al}$ and
$\bar{a}_{\dal}$ mix $y_{\al}$ with $\bar{y}_{\dal}$ via
\eqref{A-def}.

\abz
{}From \eqref{W0} and \eqref{star-prop} it follows that
\be\label{der}
\D_0\big(\vac f(A|x)\big)=\vac\Big(\hat d-\ff12dK^{AB}\ff{\p^2}{\p
A^A\p A^B}\Big)f(A|x)\,,
\ee
where the $x$--dependence of $A_A$ has been taken into account
with the aid of \eqref{glob} and \eqref{zero-W} so that the
differential $\hat d$ in \eqref{der} only accounts the
manifest $x$ dependence, {\it i.e.,} $ \hat d A=0$. Using (\ref{der}),
the HS equations that remain to be solved reduce to
\be\label{S-eq}
[s_\al, s_\gb]_*=2\gep_{\al\gb}(1+M\K)\,,
\ee
\be\label{W-eq}
\Q s_\al-[\go,s_\al]_*=0\,,
\ee
\be\label{DW-eq}
\Q\go-\go*\wedge\go=0\,,
\ee
and their complex conjugated, where
\be\label{Q}
\Q=\hat d-\ff12d\gk^{\al\al}\ff{\p^2}{\p a^\al\p a^{\al}}\,.
\ee
The operator $\Q$ has the following properties inherited from $\D_0$
\be\label{Leibtnitz}
\Q\big(f(a|x)*g(a|x)\big)=\Q f(a|x)*g(a|x)+f(a|x)* \Q g(a|x)\,,
\ee
\be
\Q^2=0\q \Q a_\al=0\q \Q \K=0\,.
\ee
Remarkably, despite  $\Q$ \eqref{Q} contains second
derivatives in the oscillators, it respects the chain rule
\eqref{Leibtnitz}. This is because the
star-product \eqref{in-star} is $x$-dependent, so that, acting on the star-product
$*$, $\hat d$ effectively compensates the terms that would spoil
\eqref{Leibtnitz}. Note that a similar construction was recently discussed in
\cite{Gelfond:2008ur} in a
different context of the definition of the ring of solutions of
unfolded HS field equations (see Appendix of \cite{Gelfond:2008ur}).

\abz Since we consider purely bosonic problem, being an even function of
$a$-oscillators, the connection \eqref{con}
commutes to the Klein operators \eqref{Klein1} and therefore
solves \eqref{DB}. (In presence of fermions this would not be true.)
In Section (\ref{detail}) we solve the equations
\eqref{S-eq}-\eqref{DW-eq} obtaining the following final result
for HS BH
\be\label{S-fin}
S_{\al}=z_\al+M\vac\ff{a^{+}_\al}{r}\int_{0}^{1}dt\, \exp{\Big(\ff
t2\gk^{-1}_{\gb\gb}a^{\gb}a^{\gb}\Big)}\,,
\ee
\be\label{bS-fin}
\bar{S}_{\dal}=\bar{z}_{\dal}+M\vac\ff{\bar{a}^{+}_{\dal}}{r}\int_{0}^{1}dt\,
\exp{\Big(\ff
t2\bar{\gk}^{-1}_{\dgb\dgb}\bar{a}^{\dgb}\bar{a}^{\dgb}\Big)}\,,
\ee
\be\label{B-fin}
B=\ff{4M}{r}\exp{\Big(\ff12\gk^{-1}_{\al\gb}y^{\al}y^{\gb}+
\ff12\bar{\gk}^{-1}_{\dal\dgb}\bar{y}^{\dal}\bar{y}^{\dgb}+
\gk^{-1}_{\al\gga}v^{\gga}{}_{\dal}y^{\al}\bar{y}^{\dal}\Big)}\,,
\ee
\be\label{W-fin}
W=W_0+\ff{M}{8r}\vac d\tau^{\al\al}a^{+}_\al
a^{+}_\al\int_{0}^{1}dt\, (1-t)\exp\Big(\ff
t2\gk^{-1}_{\gb\gb}a^{\gb}a^{\gb}\Big)+c.c.+F_K\F\,,
\ee
where
\be\label{tau}
\tau_{\al\al}\equiv\ff{\gk_{\al\al}}{r}\,,
\ee
\be\label{f0}
\F=-\ff{M}{8}(\tau_{\al\al}\go^{\al\al}+
\bar{\tau}_{\dal\dal}\bar{\go}^{\dal\dal})+\ff{M}{4r}h^{\al\dal}(v_{\al\dal}+k_{\al\dal})
\ee
and the Kerr-Schild vector $k_{\al\dal}$ is defined in \eqref{KS}.
Let us stress, that it is the Ansatz \eqref{con}, that effectively
implied holomorphic factorization of the oscillators $a_{\al}$
and $\bar{a}_{\dal}$, allowed us to integrate the equations on HS connection.
Note also, that \eqref{S-fin} and \eqref{bS-fin} do not correspond
to the standard gauge choice of the HS
perturbative analysis of \cite{Vas4-more,Vas-rev} with
$S(Z,Y)|_{Z=0}=0$. Hence, to reproduce
HS Kerr-Schild solution \eqref{Fr} at first order in $M$ one has to
apply an appropriate HS gauge transformation. In the first order, such a gauge
transformation is $W^{can}=W+\D_0 g$, where
$g=-\ff12\int_{0}^{1}dtz^{\al}S_{\al}|_{z\to tz}+c.c.+g_0(Y|x)$
with an arbitrary $g_0(Y|x)$.

\subsection{Details of analysis}
\label{detail}

HS constraints \eqref{SS}, \eqref{BS} have been  reduced to
Eq.~\eqref{S-eq}. By analogy with the standard perturbative
analysis \cite{Vas4-more,Vas-rev}, in the first order in $M$ this
gives $ \ff{\p}{\p a^\al}\gs^{\al}=M\K\,, $ that can be solved in
the form
\be\label{sigma}
\gs^{\pm}_{\al}(a|x)=\ff{M}{r}a^{\pm}_{\al}\int_{0}^{1}dt\,\exp\Big(\ff
t2\gk^{-1}_{\gb\gb}a^{\gb}a^{\gb}\Big)\q a^{\pm}_{\al}\equiv
\pi^{\pm}_{\al}{}^{\gb}a_{\gb}\,,
\ee
where $\pi^{\pm}_{\al\gb}$ are projectors \eqref{proj}. This
solves \eqref{S-eq} exactly because $ [\gs^{\pm}_{\al},
\gs^{\pm}_{\gb}]_*=0 $. Indeed, the projectors \eqref{proj} make
the antisymmetric matrix $ [\gs^{\pm}_{\al}, \gs^{\pm}_{\gb}]_*$
one-dimensional and hence zero. Choosing for definiteness the
plus sign in \eqref{sigma} we obtain \eqref{S-fin}-\eqref{B-fin}.

An important property of \eqref{sigma} is that
\be\label{Qs}
\Q\gs_{\al}=-\ff14\ff{\p}{\p a^{\al}} \gO^{\gb\gb}\{a_{\gb},
\gs_{\gb}\}_*\,,
\ee
where $\gO_{\al\al}(x)$ is the $sp(2)$ flat connection
(except may
be $r=0$, cf. \eqref{tau})
\be\label{sp2-zero}
\gO_{\al\al}=d\tau_{\al}{}^{\gga}\tau_{\gga\al}\q
d\gO_{\al\al}-\gO_{\al}{}^{\gga}\wedge\gO_{\gga\al}=0\,.
\ee

To find HS connection 1-form $W(Y,Z|x)$ corresponding to
\eqref{S-fin}-\eqref{B-fin} we start with the equation
\eqref{W-eq}. Let us solve it perturbatively. In the first order in
$M$ we have
\be
\ff{\p\go}{\p a^{\al}}=-\ff12\Q\gs_{\al}\,.
\ee
Using \eqref{Qs}, the first-order HS connection can be written
in the following remarkable form
\be\label{omega1}
\go(a|x)=\ff18\gO^{\al\al}\{a_{\al},\gs_{\al} \}_*+f_0+O(M^2)\,,
\ee
where, $f_0(x)$ is some $a_\al$--independent 1-form.

\abz The observation \eqref{Qs}, \eqref{omega1} suggests the exact
solution of \eqref{W-eq}. Indeed, one may note that the first term
in \eqref{omega1} is the linearized part of
$\ff18\gO^{\al\al}(s_\al*s_\al-a_\al*a_\al)$. Using that
bilinears of the deformed oscillators generate their $sp(2)$
rotations, \be\label{sp2} T_{\al\al}=s_{\al}*s_{\al}\q
[T_{\al\al}, s_{\gb}]_*=4\gep_{\al\gb}s_{\al}\,, \ee we obtain the
exact solution of \eqref{W-eq} in the following simple form
\be\label{W}
\go(a|x)=\ff18\gO^{\al\al}(s_{\al}*s_{\al}-a_{\al}*a_{\al})+f_0\,,\quad
\bar{\go}(\bar
a|x)=\ff18\bar{\gO}^{\dal\dal}(\bar{s}_{\dal}*\bar{s}_{\dal}-
\bar{a}_{\dal}*\bar{a}_{\dal})+\bar{f}_0\,. \ee Remarkably, the
connection \eqref{W} in fact does not contain the $O(M^2)$ terms,
{\it i.e.,} \be\label{omega2}
\go_2(a|x)=\ff18\gO^{\al\al}\gs_{\al}*\gs_\al=0\,. \ee The
simplest way to see this is to observe that from \eqref{W-eq} it
follows $ \ff{\p\go_2}{\p a^\al}=\ff12[\go_1, \gs_\al]_* $ and
hence, $ \pi^{-}_{\al}{}^{\gga}\ff{\p}{\p a^{\gga}}\go_2(a|x)=0\,,
$ so that $\go_2=\go_2(a^-|x)$. On the other hand, it is easy to
see that $\go_2(a|x)$ (\ref{omega2}) is an entire function of the
oscillators $a^\pm$ such that \be (a^+ \frac{\p}{\p a^+}
-a^-\frac{\p}{\p a^-})\go_2(a^\pm|x)=2\go_2(a^\pm|x)\,. \ee Hence,
it should be zero. Note that the straightforward verification of
this fact, that was also completed, is not trivial, implying
interesting integral identities.

Having solved \eqref{DS}, it remains to verify the HS
zero-curvature equation \eqref{DW-eq} to determine the HS
connection completely. Plugging \eqref{W} into \eqref{DW-eq} and
noting that \be d\gO^{\al\al}\{a_\al,
\gs_\al\}_*=-Md\gO^{\al\al}\tau_{\al\al} \ee one finds
\be\label{f} df_0=\ff{M}{16}d\tau^{\al\gga}\wedge
d\tau_{\gga}{}^{\al}\tau_{\al\al}\,. \ee Note, that the r.h.s. of
\eqref{f} is consistent with $d^2 f_0=0$. Indeed,
$d^2f_0=\ff{M}{16}d\tau^{\al\gga}\wedge d\tau_{\gga}{}^{\al}\wedge
d\tau_{\al\al}$ which is a $3d$ volume 3-form for the ``frame"
$E_{\al\al}=d\tau_{\al\al}$, that is however zero since
$\tau^{\al\al}\tau_{\al\al}=const$. Note that $f_0$ contributes to
the HS connection \eqref{con} via its real combination
$\F=f_0+\bar f_0$ that can be chosen in the form \eqref{f0}. It is
straightforward to verify that \eqref{f0} satisfies \eqref{f} plus
its complex conjugated. Thus, the final result for HS connection
is \eqref{W-fin}.

\section{Symmetries}\label{global}
Global symmetries of a $4d$ static BH include $SO(3)$ spatial
rotations and  $R^1$  time translations both in asymptotically
Minkowski  and in  $AdS_4$ (in fact, its universal covering)
geometry. Infinitesimally, they form  algebra $su(2)\oplus gl(1)$.
A static Reissner--Nordstr\"{o}m BH is in addition characterized
by its electric charge $e$, reproducing Schwarzschild BH at $e=0$.
The critical value of charge $ e^2 = M^2, $ that corresponds to
the extremal BH, is characterized by  the coincidence of two BH
horizons \cite{Wald} and by SUSY, being BPS \cite{GH82}. Note
that, at the free field level, the $s=1$ field in \eqref{B-fin} is
just the Maxwell field strength of the $AdS_4$
Reissner--Nordstr\"{o}m potential.

In this section we  analyze  global symmetries of the obtained
static HS BH solution, showing in particular that (i) its
space-time  symmetry is $su(2)\oplus gl(1)$, (ii) it is
supersymmetric, preserving a quarter of $4d$ $\mathcal{N}=2$ SUSY
of the nonlinear HS model of \cite{Vas4-more}, and (iii)
it possesses infinite dimensional HS extension of
 (super)symmetries of (i) and (ii).

\subsection{Bosonic symmetries}
{}From \eqref{gauge-B} it follows that the leftover global
symmetry parameter $\gep_{0}(Y|x)$ should satisfy $ \gep_0\*
B-B\*\tilde{\gep}_0=0\,. $ (Note that all symmetries with
$Z$-dependent parameters $\gep_{0}(Z,Y|x)$ are spontaneously
broken because of the $Z$-dependent vacuum part of $S$.) Taking
into account \eqref{B}, this gives \be\label{HS-glob1} \gep_0\*
F_K-F_K\*\gep_0=0\,. \ee Since the Fock vacuum  $F_K$ satisfies
\eqref{fock}, the general solution  of \eqref{HS-glob1} is
\be\label{BH-glob} \gep_0(Y|x)=\sum_{m,n=1}^{\infty}f_{0 A(m),
B(n)}(x)\underbrace{Y_{+}^{A}\*\dots\*
Y_{+}^{A}}_{m}\*\underbrace{Y_{-}^{B}\*\dots\*
Y_{-}^{B}}_{n}+c_0(x)\,. \ee Now we observe that any $\gep_0(Y|x)$
\eqref{BH-glob} commutes to any $F_K\phi(A|x)$ as one can  see
from Eq.~(\ref{def}). As a result, the only nontrivial condition
in \eqref{gauge-B} that remains is $
d\gep_0+[\gep_0, W_0]_\*=0\,.\quad 
$
This requires $\gep_0(Y|x)$ (\ref{BH-glob}) ({\it i.e.,}
 $f_{0 A(m), B(n)}(x)$ and $c_0(x)$) be $AdS_4$
covariantly constant.

A maximal finite dimensional subalgebra of \eqref{BH-glob} is
spanned by the bilinears of $Y_{-}$ and $Y_+$ and constants. In
particular, it contains generators of $su(2)\oplus gl(1)$
\be\label{low-sym}
T^{AB}=Y_{+}^{(A}Y_{-}^{B)}\,,\qquad T=Y_{-A}Y_{+}^{A}
\ee
that belong to the class (\ref{BH-glob}). Since the algebra
$sp(4)$ of $AdS_4$ space-time symmetries is spanned by  various
bilinears of $Y_A$, this $su(2)\oplus gl(1)$
describes  space-time symmetries that remain unbroken. Hence, the obtained
solution indeed describes a spherically symmetric static BH. Note that the
$Y$-independent constant parameter in \eqref{BH-glob} describes a
$u(1)$ inner symmetry. The full set
of parameters (\ref{BH-glob}) describes an infinite dimensional HS
algebra of global symmetries of the HS BH solution.

\subsection{Supersymmetry}

The solution \eqref{S-fin}-\eqref{W-fin} turns out to be
supersymmetric, which is most easily seen from the embedding of
the bosonic HS equations considered so far into the
$\mathcal{N}=2$ supesymmetric nonlinear HS system of
\cite{Vas4-more} (see also \cite{Vas-rev}) which has the form
\be\label{ww}
d\mathcal{W}-W\*\wedge \mathcal{W}=0\q d\mathcal{B}-[\mathcal{W},
\mathcal{B}]_\*=0\q d\mathcal{S}-[\mathcal{W},\mathcal{S}]_\*=0\,,
\ee
\be
\label{ss} \mathcal{S}\*\mathcal{S}=dz_\al
dz^\al(1+\mathcal{B}\*kv)+d\bar{z}_{\dal}d\bar{z}^{\dal}(1+\mathcal{B}\*\bar
k\bar v)\q [\mathcal{S},\mathcal{B}]_\*=0\,,
\ee
where $\mathcal{W}=\mathcal{W}(Z,Y; k, \bar k|x)$,
$\mathcal{B}=\mathcal{B}(Z,Y; k, \bar k|x)$
and $k, \bar k$ are the exterior Klein operators that satisfy
$k^2=\bar k^2=1$, $[k, \bar k]=[k,
dx^{\mu}]=[\bar k, dx^{\mu}]=0$ and
\be
\label{kcom}
k f(Z, Y; dZ; k, \bar k|x)=f(\tilde{Z}, \tilde Y; d\tilde{Z}; k,
\bar k|x)k\,,\quad \bar k f(Z,Y; dZ; k, \bar k|x)=f(-\tilde Z,
-\tilde Y; -d\tilde{Z}; k, \bar k|x)\bar k\,,
\ee
where $\tilde U_A=(-u_\al, \bar{u}_{\dal})$ for $U_A=(u_\al, u_{\dal})$.

It is important to note that the system (\ref{ww}), (\ref{ss}) describes
the doubled set of massless fields where all fields of integer and half-integer
spins appear in two copies. Its bosonic sector consists of two independent
subsystems described by the equations (\ref{DW})-(\ref{BS}), which can be projected
from the system (\ref{ww}), (\ref{ss}) by the projectors
\be
P^{\pm}=\frac12 (1\pm k\bar k)\,,\qquad
P^{\pm}P^{\pm}=P^{\pm}\,,\qquad P^{\mp}P^{\pm}=0\,,
\ee
that by (\ref{kcom}) commute to boson fields, that are even in
spinorial variables, but not to fermions. As such, the bosonic
reduction of the system (\ref{ww}), (\ref{ss}) describes two
parallel bosonic worlds. Each of them is not
 supersymmetric because $P^\pm$ do not commute to fermions.
Conventional $\mathcal{N}=2$ SUSY
is achieved in the ``diagonal world" described by the metric $g^{nm}=\half(g^{+nm} +g^{-nm})$.

A natural BH solution of the supersymmetric HS theory is a
combination of BH solutions in each of the bosonic sectors, where,
{\it a priori}, one can choose solutions with unrelated
$K^\pm_{AB}$. We wish however to consider the case where
$K^+_{AB}=- K^-_{AB}$ and the corresponding Fock vacua (\ref{exp})
$F_K^\pm$ have opposite properties\footnote{Let us note that
although the Fock vacua $F_K^\pm$ have ill-defined (infinite)
mutual star--product this does not matter because, living in
``parallel worlds" ({\it i.e.,} being multiplied by $P^\pm$), they
never meet.} (\ref{fock}) \be Y_{\mp A} F_K^\pm = F_K^\pm Y_{\pm
A} =0\q F^\pm_K= \exp{\Big (\pm \half K_{AB}Y^AY^B}\Big )\,. \ee

Consider the bosonic solution of the $\mathcal{N}=2$ supersymmetric HS system
of the form
\be\label{embd}
\mathcal{W}=(P^+W^+ + P^-W^-)\,,\quad \mathcal{B}=\half
(B^+(k+\bar k) + i B^-(k-\bar k))\,,\quad \mathcal{S}= (P^+S^+ +
P^-S^-)\,.
\ee
 (Note that  the factor of
$i$ in the definition of $\mathcal{B}$ in Eq.(\ref{ss}) is enforced by the reality
conditions that conjugate $k$ and $\bar k$ \cite{Vas4-more}.)
The labels $\pm$ refer to the solutions in the sectors of $P^{\pm}$
that may have different masses $M^\pm$.
It is however convenient to demand
the vacuum fields to coincide: $W^\pm_0=W_0$, $S^\pm_0=S_0$. In the  case
where $M^+$ or $M^-$ vanishes, the respective world is  $AdS_4$.

A global symmetry parameter $\gep(Y; k \bar k|x)$ now
depends on $k\bar k=P^+-P^-$ and should satisfy the conditions
\be\label{N2}
[\mathcal{B}, \gep]_\*=0\,,\qquad [\gep,
\mathcal{S}]_\*=0\,,\qquad d\gep-[\mathcal{W}, \gep]_\*=0\,.
\ee
These are verified by $AdS_4$ covariantly constant $\gep$
that admit the following two representations
\be
\label{susy}
\gep(Y ;k\bar k |x) = \gep^+_{lA} (Y)\star Y^A_- P^+ +\gep^-_{lA}(Y) \star Y^A_+ P^-+c_0=
P^+ Y^A_+ \star\gep^+_{rA}(Y) + P^- Y^A_- \star\gep^-_{rA}(Y) +c_0
\ee
with some $\gep^\pm_{lA} (Y)$ and $\gep^\pm_{rA} (Y)$. Indeed, for instance, the terms
proportional to $F^+ P^+$ are annihilated by $Y_-$ from the left, $Y_+$ from
the right and $P^-$ from the both sides.

Clearly, the
bosonic global symmetry parameters (\ref{BH-glob}) belong to the class (\ref{susy}).
However, now the parameters that  are odd in $Y_A$ are
also allowed. In particular, global SUSY
with an $AdS_4$ covariantly constant spinor parameter $\gep_-^A(x)$
\be
\gep(Y ;k\bar k |x) = \gep_-^A(x)P^- Y_{-A}= \gep_-^A(x) Y_{-A} P^+\q D_0 \gep_-^A(x)=0\q
\Pi_+^A{}_B(x) \gep_-^B(x)=0
\ee
belongs to this class (recall that $Y_A$ anticommute to $k\bar
k$). This global SUSY is a quarter of the $\mathcal{N}=2$ SUSY
with the supergenerators $Q^1_A=Y_A$ and $Q^2_{A}=ik\bar k Y_A$
\cite{FVau} of the $AdS_4$ vacuum of the system (\ref{ww}),
(\ref{ss}). Hence, the HS BH subalgebra corresponds to a
$\frac{1}{4}$ BPS state. This strongly indicates that the obtained
HS BH solution should be extremal.

Covariantly constant solutions of (\ref{susy})
form an infinite dimensional HS superalgebra of global symmetries
of the obtained BH solution.

\section{Conclusion}
\label{conc}

The new exact solution of $4d$ bosonic HS theory  announced in
this paper is a HS generalization of a static BH in GR. At
the  free field level it contains Schwarzschild BH in the spin two
sector.  However, the contribution of HS fields is important in
the strong field regime as is demonstrated by the remarkable fact that
the nonlinear corrections cancel out for static BH in the full HS
theory, reducing HS nonlinear equations to the free ones. This
property is a HS analogue of the fact that the usual Kerr--Schild
Ansatz reduces nonlinear Einstein equations to free Pauli-Fierz
ones for a $4d$ BH. Let us stress that massless HS fields do not
satisfy HS field equations in the BH background  unless the
HS interactions are switched on via the terms bilinear in the HS connections
in the nonAbelian HS curvatures on the left hand side of (\ref{DW}).

\abz In the proposed construction, static HS BH is described in
terms of a Fock vacuum in the star--product algebra in the
auxiliary twistor space. This Ansatz effectively projects $4d$
HS equations to $3d$ HS equations of \cite{ProkVas} that describe
$3d$ massive matter fields and can be solved with the aid of
Wigner's deformed oscillators. The BH mass $M$ coincides with the
vacuum value $B_0=\nu$ of \cite{ProkVas} that sets the mass scale
of $3d$ interacting massive fields. This reduction suggests an
interesting duality between $AdS_3$ massive fields of a mass scale
$\nu$ and $4d$ HS BH of mass $M$, $ \nu=\gl GM, $ where $-\gl^2$
is the cosmological constant and $G$ is the Newton constant. More
generally, the obtained results indicate that near BH fluctuations
in HS theory describe a HS theory in the lower dimension thus
providing a nontrivial dimensional compactification mechanism
analogous to the brane picture in String Theory.

At the linearized level, where other fields do not contribute to
the metric, the obtained solution reproduces $AdS_4$ Schwarzschild
BH in the $s=2$ similarly to the charged
Reissner--Nordstr\"{o}m BH that involves second order contribution of
the electromagnetic field via its stress tensor. That the obtained
solution exhibits leftover SUSY strongly indicates that it should
be extremal. We believe that it can be further generalized to the
NUT case using the phase ambiguity of $4d$ nonlinear HS equations
of \cite{Vas4-more} as well as to the Kerr BH,
extremal and not. More generally, a
generalization to $d$ dimensions is a challenging issue both at the free and at the
nonlinear level. The preliminary analysis indicates that our
method should work in various dimensions. It looks especially
promising in the context of black rings \cite{bring} that exist in
$d\geq 5$.

\abz An exciting problem for the future study is to explore
physical properties of the obtained solution. Here, the key
question is how do  small fluctuations propagate in the
HS BH  background? Its analysis should  shed light
on such fundamental concepts in HS BH physics as
 horizons, trapped surfaces {\it etc}. Since HS theory is essentially nonlocal
 at the interaction level,
 involving higher  derivatives for higher spins, the analysis of
 these issues should be done by new means beyond standard GR machinery.
 For example, in HS theory it is not granted that the geodesic motion has
 much to do with signal propagation in the strong field regime. Moreover, in the HS theory
with unbroken HS symmetries it is not even clear how to define a
metric tensor beyond the linearized approximation. The study of
thermodynamical interpretation of the HS BH is also of primary
importance. We plan to consider these problems in the future.

\section*{Acknowledgement}
We acknowledge participation of  A.S. Matveev at the early stage
of this work and would like to thank him for many useful
discussions. Also we are grateful to C. Iazeolla, A.G. Smirnov,
and P. Sundell for useful comments and to A.A. Tseytlin for
communication to us some relevant references. This research was
supported in part by, RFBR Grant No 08-02-00963, LSS No 1615.2008.2
and Alexander von Humboldt Foundation Grant PHYS0167.

\renewcommand{\theequation}{A.\arabic{equation}}
\renewcommand{\thesubsection}{A.}
\renewcommand{\thesection}{Appendix. Conventions}

\section{}


For any Lorentz vector $\xi^n$, its two-component
spinor counterpart is
$
\xi_{\al\dal}=\xi^n\gs_{n,\al\dal}\,,
$
where $\gs_{n,\al\dal}=(I_{\al\dal}, \gs_{i,\al\dal})$ contains unity matrix $I$
along with Pauli matrices. Latin indices are raised
and lowered by Minkowski metric $\eta_{mn}$.
Spinorial indices are raised and lowered according to
\be
A_{\al}=A^{\gb}\gep_{\gb\al}\q A^{\al}=A_{\gb}\gep^{\al\gb}\q
\bar{A}_{\dal}=\bar{A}^{\dgb}\gep_{\dgb\dal}\q
\bar{A}^{\dal}=\bar{A}_{\dgb}\gep^{\dal\dgb}\,.
\ee
$Sp(4)$ indices $A=(\al,\dal)=1\dots 4$ are raised and lowered
by the canonical symplectic form
\be
\gep_{AB}=-\gep_{BA}=\left(
\begin{array}{cc}
\gep_{\al\gb} & 0\\
0 & \gep_{\dal\dgb}\\
\end{array}
\right)\q A_{A}=A^B\gep_{BA}\q A^A=A_B\gep^{AB}\,.
\ee

\abz To distinguish between two types of projectors $\Pi_{\pm AB}$ and
$\pi^{\pm}_{\al\gb}$ we use the convention with lower
and upper labels $\pm$
assigned to the objects projected by $\Pi_{\pm AB}$ and
by $\pi^{\pm}_{\al\gb}$, respectively. For example,
\be
Y_{\pm A}=\Pi_{\pm A}{}^{B}Y_B\,,\quad
y_{\pm\al}=\Pi_{\pm\al}{}^{B}Y_B\,,\quad
\bar{y}_{\pm\dal}=\Pi_{\pm\dal}{}^{B}Y_B\,,
\ee
but
\be
y^{\pm}_{\al}=\pi^{\pm}_{\al}{}^{\gb}y_{\gb}\q
\bar{y}^{\pm}_{\dal}=\bar{\pi}^{\pm}_{\dal}{}^{\dgb}\bar{y}_{\dgb}\,.
\ee

\end{document}